# Resolution-enhanced OCT and expanded framework of information capacity and resolution in coherent imaging

Nichaluk Leartprapun[1] and Steven G. Adie[1,*]

[1]Nancy E. and Peter C. Meinig School of Biomedical Engineering, Cornell University, Ithaca, New York 14853, USA
*Correspondence should be addressed to sga42@cornell.edu

**Abstract**
Spatial resolution in optical microscopy has traditionally been treated as a fixed parameter of the optical system. Here, we present an approach to enhance transverse resolution in beam-scanned optical coherence tomography (OCT) beyond its aberration-free resolution limit, without any modification to the optical system. Based on the theorem of invariance of information capacity, resolution-enhanced (RE)-OCT navigates the exchange of information between resolution and signal-to-noise ratio (SNR) by exploiting efficient noise suppression via coherent averaging and a simple computational bandwidth expansion procedure. We demonstrate a resolution enhancement of 1.5× relative to the aberration-free limit while maintaining comparable SNR in silicone phantom. We show that RE-OCT can significantly enhance the visualization of fine microstructural features in collagen gel and *ex vivo* mouse brain. Beyond RE-OCT, our analysis in the spatial-frequency domain leads to an expanded framework of information capacity and resolution in coherent imaging that contributes new implications to the theory of coherent imaging. RE-OCT can be readily implemented on most OCT systems worldwide, immediately unlocking information that is beyond their current imaging capabilities, and so has the potential for widespread impact in the numerous areas in which OCT is utilized, including the basic sciences and translational medicine.

**Introduction**

The enhancement of resolution[1-5] has been an important and on-going pursuit in all fields of imaging, including coherent[6-27] and incoherent[28-38] optical microscopy. One approach for resolution enhancement is spatial-frequency bandwidth expansion (i.e., increasing the numerical-aperture (NA)) via aperture synthesis. This family of techniques utilizes multiple measurements of the sample that provide access to different spatial frequencies beyond the bandwidth support of a single measurement. A well-known technique in incoherent microscopy is structured illumination microscopy (SIM), which uses different illumination patterns to shift the spatial frequency coverage of the optical system[32-34]. In coherent microscopy, the bandwidth support of the optical system can be shifted via use of multiple illumination angles, as has been implemented with off-axis holography[6-8] and Fourier ptychography[9-11] (which performs incoherent imaging at different illumination angles, combined with phase retrieval methods to reconstruct the complex optical field). In optical coherence tomography (OCT), interferometric synthetic aperture microscopy (ISAM) utilizes synthetic aperture methods to overcome the trade-off between resolution and depth-of-field, in order to reconstruct depth-invariant focal-plane resolution from a single volumetric measurement with the optical focus at a fixed depth[12,13]. Combining full-field OCT with holography, holoscopy can similarly achieve focal-plane resolution across all depths[14]. More recently, optical coherence refraction tomography (OCRT) utilizes sample rotation combined with an (incoherent) Fourier synthesis technique reminiscent of X-ray computed tomography to effectively 'replace' the lateral resolution of a low-NA imaging beam by the superior axial resolution of OCT[15].

Another family of techniques aims to correct aberrations in order to restore ideal focal-plane resolution. Hardware-based adaptive optics (HAO) has been implemented in coherent[16-19] and incoherent[28-31] microscopy. In coherent imaging, access to the complex optical field can enable computational aberration correction post-data-acquisition. Computational adaptive optics (CAO) modifies the pupil phase of complex OCT tomograms to correct both defocus and optical aberrations, in order to restore aberration-free focal-plane resolution[20-27]. Indeed, reaching the ideal aberration-free resolution supported by the optical system (i.e., without entering the super-resolution regime), especially in complex media such as biological samples, is the aim of many adaptive optics or computational aberration correction methods in optical microscopy.

Spatial resolution (or equivalently, spatial-frequency bandwidth) in optical microscopy has traditionally been treated as a *fixed* parameter for a given optical system. However, the *theorem of invariance of information capacity* suggests that resolution of an optical system is a *tuneable* parameter that can be flexibly modified *without* altering the optical system[3-5]. Cox and Sheppard described the information capacity of an optical system as the product of its space-bandwidth products (SBP) over all spatial dimensions, time-bandwidth product (TBP), and its signal-to-noise ratio (SNR) on the logarithmic scale[4]. The theorem of invariance of information capacity states that it is not the spatial-frequency bandwidth (and therefore resolution), but the information capacity of an optical system that is invariant[3-5]. It follows that spatial resolution can, in theory, be enhanced beyond the aberration-free limit of a given optical system through an exchange of information between spatial-frequency bandwidth and SNR, while keeping the information capacity constant.

The framework of informational capacity and its invariance underscores several unique advantages of coherent over incoherent imaging. A coherent imaging system inherently supports twice the information capacity of an equivalent incoherent system because each pixel is described by both magnitude and phase of the complex optical field, as opposed to a single intensity value[4,5,39]. Using quantum Fisher information formalism, others have also shown that the resolution limit in traditional intensity-based imaging techniques can be overcome by making use of phase information[40-42]. Bilenca et al. derived the scattering limit to the information capacity of depth-resolved coherent imaging (using OCT as a case study) through turbid media as a function of SNR[43]. Furthermore, coherent averaging of complex tomograms provides a more efficient method for noise suppression than incoherent (magnitude-only) averaging in OCT, due to the decorrelation of random phase noise[44-46]. Coherent averaging has also been used for multiple scattering suppression[47-49]. The efficient noise suppression via coherent averaging presents an opportunity to expand the information capacity of a coherent imaging system via the enhancement of SNR.

Resolution-enhanced (RE)-OCT is a spatial-frequency bandwidth expansion approach that computationally enhances transverse spatial frequencies beyond the traditional bandwidth support (i.e., beyond the aberration-free resolution limit) of a beam-scanned OCT system. Unlike existing aperture synthesis techniques[6-11,15,32-34], it does not rely on diversity in the illumination schemes to access traditionally unseen spatial frequencies, and so can readily be implemented on existing OCT systems. RE-OCT is based upon the premise of information exchange between spatial-frequency bandwidth and SNR, as governed by the theorem of invariance of information capacity[3-5]. It navigates this exchange of information by exploiting efficient noise suppression via coherent averaging and a simple computational bandwidth expansion procedure. RE-OCT harnesses the benefit of coherent averaging to (for the first time) enhance resolution in OCT.

In this paper we first discuss the underlying principle of RE-OCT based on the information capacity framework. Then, we demonstrate noise suppression via coherent averaging and analyse its impact on the OCT signal in not only the space, but also the spatial-frequency domain. Our analysis of the impact of coherent averaging in the transverse spatial-frequency domain provides a new perspective compared to prior work in OCT (which has only investigated the impact of coherent and incoherent averaging in the space domain[44-46]). Next, we demonstrate resolution enhancement by RE-OCT in both resolution phantom and biological samples (collagen gel and *ex vivo* mouse brain). We then analyse the factors that limit resolution enhancement in RE-OCT by comparing experimental results to simulations. Lastly, we leverage insights from our analysis in the spatial-frequency domain to present an expanded framework of information capacity and resolution in coherent imaging. RE-OCT has the potential to have widespread impact since it can be readily implemented on most OCT systems without requiring any redesign of the optical system or specialized computationally-intensive algorithms.

## Results
**Underlying principle of resolution-enhanced OCT**
RE-OCT is a spatial-frequency bandwidth expansion approach based on the exchange of information between resolution and SNR that is supported by the underlying theorem of invariance of information capacity[3-5]. We conjecture that the transverse resolution of a tomogram acquired by a beam-scanned OCT system can be enhanced via computational bandwidth expansion (BE), where signal at higher spatial frequencies is raised via multiplication by a magnitude mask in the spatial-frequency domain (i.e., magnitude-based deconvolution). This is feasible due to the under-filling of the objective aperture (i.e., a physical bandwidth limit) that is implemented in the ubiquitous telecentric scanning scheme (Supplementary Fig. 2b) used by most OCT systems worldwide. However, deconvolution inherently amplifies noise and degrades the SNR of the image. To address this problem, we also conjecture that the SNR penalty associated with computational BE can be compensated by first suppressing the system noise—one efficient way in OCT is by coherently averaging multiple successively acquired complex tomograms[44-46]. The underlying principle of RE-OCT is illustrated in Fig. 1. RE-OCT utilizes coherent averaging for efficient noise suppression to 'earn' SNR, which can be used to 'purchase' resolution via computational BE.

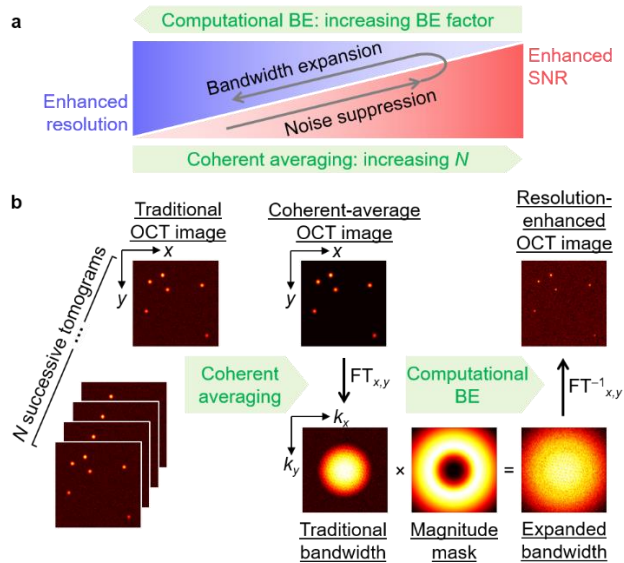

**Figure 1 Underlying principle of RE-OCT. a**, Information exchange between resolution and SNR facilitated by noise suppression and spatial-frequency bandwidth expansion. RE-OCT utilizes coherent averaging to efficiently suppress noise and enhance SNR, then SNR is sacrificed in the computational BE procedure to enhance resolution. **b**, Illustration of RE-OCT process to implement the principle in **a**. Supplementary Section IV provides a full description of RE-OCT reconstruction procedure. BE, bandwidth expansion. FT, Fourier transform.

To understand the underlying principle of RE-OCT in the context of the theorem of invariance of information capacity, we restate Cox and Sheppard's expression for the information capacity of an optical system[4] (Eq. 1).

$$C = (2L_x B_x + 1)(2L_y B_y + 1)(2L_z B_z + 1)(2TB_T + 1)\log_2(1 + s/n), \tag{1}$$

where $L$, $T$, and $B$ denote the spatial field-of-view (FOV), temporal duration, and bandwidth in the associated dimension, respectively. The first three terms represent the SBP along the three spatial dimensions while the fourth term represents the TBP. The last term represents the SNR (in bits), where $s$ and $n$ denote the average signal power and the additive noise power, respectively. Coherent averaging can be considered as the exchange of the temporal information (of an object that is *invariant* in time) for enhanced SNR. This effectively expands the information capacity of coherent-averaged versus the single-shot tomograms via the increase in the SNR term, which can be sacrificed to expand each of the transverse spatial-frequency bandwidths, $B_x$ and $B_y$. (For a full analysis of this process, see Supplementary Section I.) One scenario is to expand the bandwidth equally along each transverse dimension by a factor equal to the square root of the gain in SNR in order to, in principle, enhance resolution in the transverse plane by the same factor, without suffering any SNR penalty relative to the traditional single-shot tomogram (Supplementary Fig. 1). However, RE-OCT is not limited to this scenario; the trade-off between resolution and SNR can be flexibly navigated by tuning the number of tomograms averaged and the BE factor applied (i.e., more SNR than the amount 'earned' from coherent averaging can be sacrificed to prioritize further resolution enhancement) (Fig. 1a).

**Coherent-average noise suppression in space and spatial-frequency domains**
Noise suppression via coherent and incoherent average over $N = 1$ through 100 acquisitions was investigated in both space and spatial-frequency domain in a silicone phantom containing scattering particles (Fig. 2). Supplementary Section III discusses the effects on the image signal in the space domain, where a coherent average demonstrated a factor of $\sqrt{N}$ superior noise reduction efficiency over an incoherent average (Fig. 2a,b). Our results are consistent with previous work in OCT and theoretical trends[45,46] (see Supplementary Section III for caveats of experimentally achieving the theoretical performance).

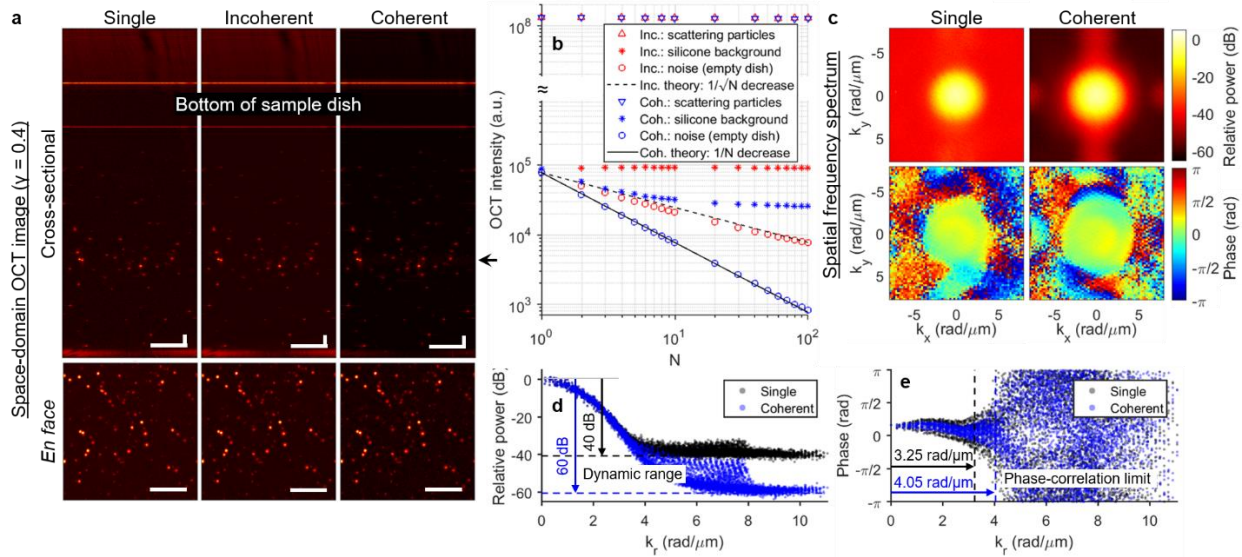

**Figure 2 Coherent-average noise suppression with 100 acquisitions in silicone phantom. a**, Single-shot, incoherent- and coherent-average OCT images (same colormap range). *En face* images correspond to focal plane, indicated by arrow. Scale bar, 50 µm. **b**, OCT intensity of scattering particles, silicone background, and noise as a function of $N$ for incoherent (red) and coherent (blue) average. Noise intensity was obtained from the "noise image" of an empty sample dish (see Methods), at the same pixel depth as the focal plane of the phantom images. **c**, Power and phase of single-shot and coherent-average images in transverse spatial-frequency domain (with spatial frequencies $k_x$ and $k_y$). Phase spectrum was obtained from the Fourier transform of a windowed region around a single particle. **d**, Power spectrum in c plotted as a function of radial transverse spatial frequency $k_r$. Dotted line indicates noise floor obtained from power spectrum of the noise image. **e**, Phase spectrum in c plotted as a function of $k_r$. Dotted line

indicates $k_r$ limit beyond which phase becomes decorrelated (local standard deviation > 0.2 rad, see Methods). Supplementary Section VII investigates relations between $N$, DR and phase-correlation limit. Calculations of OCT intensity, DR, and phase-correlation limit are described in Methods.

Here, we additionally investigated coherent-average noise suppression in the spatial-frequency domain (Fig. 2c–e), where the image is a superposition of the detected backscattered signal with a Gaussian magnitude spectrum (when imaged with a Gaussian beam) and the system noise with a uniform magnitude spectrum (circular Gaussian random variable in space). The power spectrum exhibited a dynamic range (DR) in the spatial-frequency domain, as measured from the power at DC to the noise floor (Fig. 2d). A coherent average over $N = 100$ acquisitions resulted in a suppressed noise floor that led to an increase in DR of 20 dB, corresponding to noise reduction by a factor of 100 (Fig. 2d). In other words, the coherent average has revealed phase-stable but low-magnitude backscattered signal at higher spatial frequencies, which were originally below the noise floor in the single-shot image but are now above the suppressed noise floor. Consequently, signal phase remains correlated (i.e., absence of random phase variation) across spatial frequencies corresponding to a larger bandwidth (Fig. 2c,e). The improved correlation of signal phase after reduction in the noise floor can also be understood by considering the impact of SNR on phase noise in phase-sensitive OCT[50,51]. Supplementary Section VII investigates (in simulation) the relations between $N$, DR and phase-correlation limit in the spatial-frequency domain.

**Resolution-enhanced OCT in silicone phantom**
Resolution enhancement in silicone phantom using coherent average over 100 acquisitions and a computational BE expansion factor of 2.4× is shown in Fig. 3. (See Methods and Supplementary Section IV for a complete description of the RE-OCT reconstruction procedure.) RE-OCT achieved a RE factor of 1.5×, from the traditional aberration-free resolution of 2.1 µm to an enhanced resolution of 1.4 µm (Fig. 3a,b), while the peak signal-to-background ratio (SBR) marginally decreased from 50 dB to 48 dB (Fig. 3a,c). However, when computational BE was performed on the single-shot image, not only was the SBR substantially decreased by 10 dB, but the quality of the point spread function (PSF) also suffered (Fig. 3a,c). This penalty is a result of computational BE indiscriminately amplifying both the backscattered signal and the system noise that dominates at higher spatial frequencies (Fig. 2d,e). Resolution improved at the cost of degraded SBR as a larger BE factor was applied (Fig. 3d,e and Supplementary Movie 1). Noise suppression prior to computational BE was essential in maintaining adequate SBR as well as the quality of the PSF in RE-OCT. However, even with a coherent average over 100 acquisitions, the best achievable resolution from this experiment was limited to 1.4 µm at BE factor of 2.4; applying a larger BE factor only resulted in lower SBR and degraded PSF quality, without further improvement in resolution.

In order to investigate the factors that limit the experimentally achievable resolution enhancement, we performed the RE-OCT procedure on 6 simulated *en face* planes (3 conditions, each with and without aberrations) with properties representative of the silicone phantom images (Fig. 3d–g). (See Supplementary Section VI for information on the simulated *en face* planes.) Based on Cox and Sheppard's information capacity framework[4], the achieved RE factor is expected to be equivalent to the applied BE factor (see Supplementary Section VIII). This relationship holds true for the simulated ideal noise-free condition, in which the RE-OCT efficiency (defined as the ratio RE factor/BE factor) remained 1 up to the Nyquist limit (Fig. 3f, noise-free limit). However, the presence of system noise decreased the RE-OCT efficiency with increasing BE factor (Fig. 3f, noise only). The trends as a function of BE factor for resolution, SBR, and RE-OCT efficiency for the noise-only condition are remarkably consistent with the experimental results (Fig. 3d–f). Furthermore, the presence of scattering signal from the silicone background, in addition to system noise, caused a slight decrease in RE-OCT efficiency relative to the noise-only condition (Fig. 3f, noise with background). These results suggest that system noise is the primary limiting factor in RE-OCT. Indeed, both experiment and simulation showed that superior resolution was achieved with coherent average over larger $N$ (i.e., more noise suppression) for a BE factor of 2.4 (Fig. 3g). In addition, optical aberrations degraded resolution, SBR, and RE-OCT efficiency for all simulated conditions (Fig. 3d–g, red). As expected, the simulated condition incorporating all three contributions: system noise, silicone background, and aberrations, most closely matched the experiment.

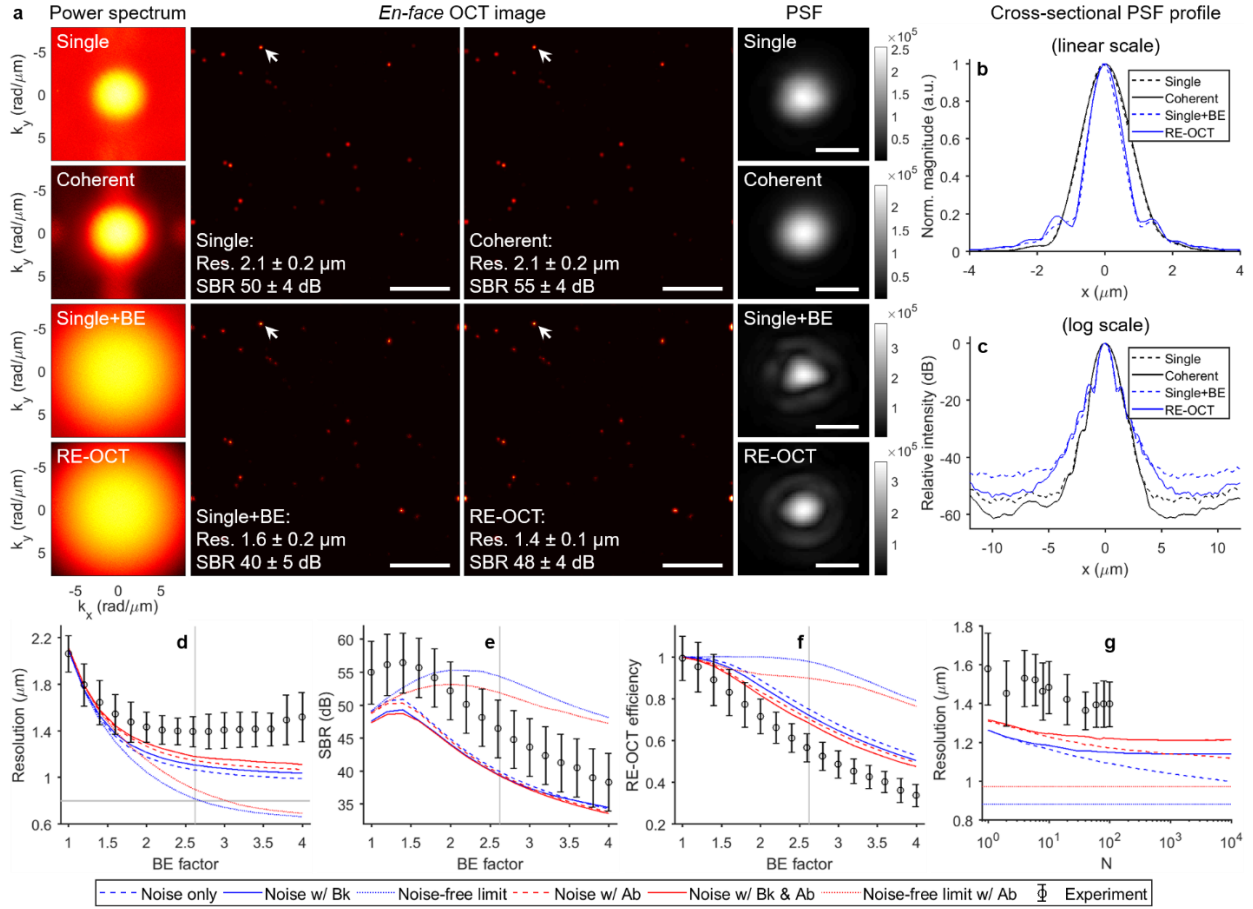

**Figure 3 RE-OCT with 100 acquisitions and 2.4× bandwidth expansion (BE) in silicone phantom. a**, Single-shot, coherent-average, BE single-shot, and RE-OCT power spectrums (log scale) and space-domain *en face* OCT image with zoomed PSF (linear scale). Resolution and SBR represent mean ± standard deviation of measurements from 11 particles. Scale bars, 40 µm (*en face* image) and 2 µm (zoomed PSF). **b** and **c**, Cross-sectional profiles of zoomed PSF in a on peak-normalized linear and log scales. **d**–**f**, Resolution, SBR, and RE-OCT efficiency as a function of BE factor from experiment and simulations for $N = 100$ of noise-free (dotted), noise only (dashed), and noise with background (solid) conditions, with (red) and without (blue) optical aberrations. Grey vertical lines indicate the Nyquist limit based on spatial sampling of 0.4 µm/pixel. **g**, Resolution as a function of *N* from experiment and simulations for BE factor of 2.4. Resolution and SBR in **a** and data point and error bars for 'Experiment' in **d**–**g** represent mean ± standard deviation of measurements from 11 particles (see Methods). Bk, background. Ab, aberrations. See Supplementary Movie 1 for a movie of **a**–**c** as increasing BE factor is applied.

### Resolution-enhanced OCT in biological samples

We implemented RE-OCT in collagen gel and *ex vivo* mouse brain, and show the best RE-OCT performances that were achieved, corresponding to BE factor of 2.0 (Figs. 4 and 5). In fibrous collagen gel, RE-OCT enhanced the visualization of the collagen fibre architecture by not only narrowing the width of the collagen fibres, but also increasing the peak signal magnitude of each fibre as a result of the resolution enhancement (Fig. 4a,b). Remarkably, low-contrast fine microstructural features, which were not clearly discernible in the traditional single-shot image due to weak signal, are more apparent in the RE-OCT image owing to the improved localization of signal energy in space (Fig. 4a, yellow arrows). In the BE single-shot image, the narrowing of fibre width can still be observed to a certain extent, but the peak signal magnitude did not improve as much (Fig. 4b). Furthermore, the SBR was degraded more severely in the BE single-shot image due to the amplification of noise without prior noise suppression (Fig. 4c). Computational BE with BE factors larger than 2.0 resulted in degraded SBR without further narrowing of the fibre width or improvement to the peak signal magnitude (Supplementary Movie 2), similar the degradation observed with BE factors larger than 2.4 in the silicone phantom (Supplementary Movie 1).

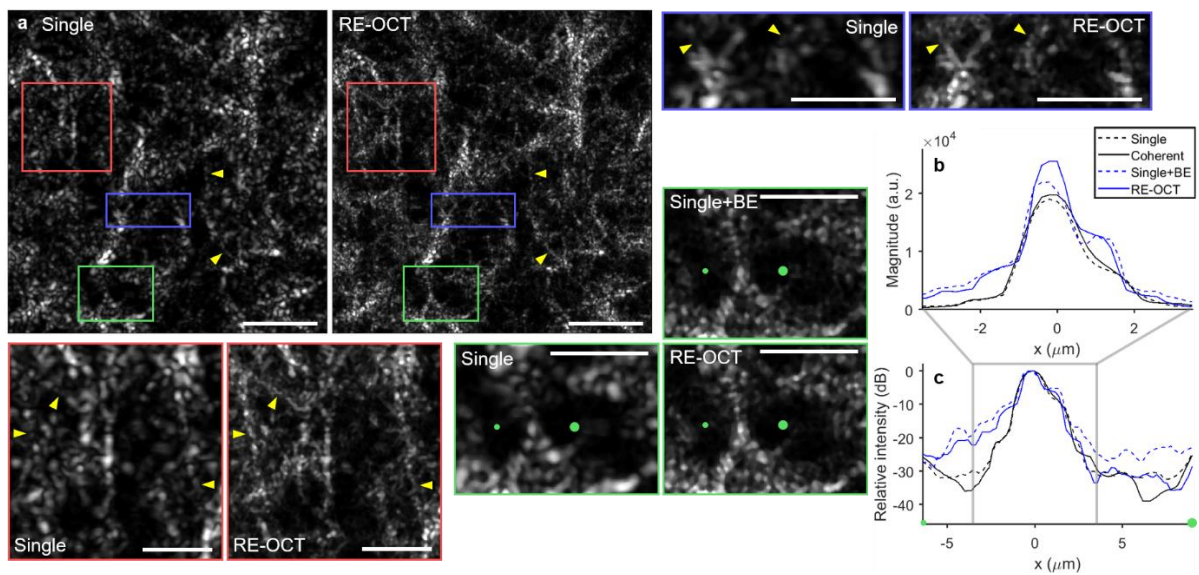

**Figure 4 RE-OCT with 100 acquisitions and 2.0× BE in fibrous collagen gel. a**, Single-shot and RE-OCT *en face* OCT images with zoomed insets regions indicated by boxes. Yellow arrows indicate fine fibre structures that can be more clearly visualized with RE-OCT. Scale bars, 40 µm (full) and 20 µm (zoomed). **b** and **c**, Cross-sectional profiles of a line connecting from small to larger green dots in the green zoomed insets in a on linear and peak-normalized log scales. See Supplementary Movie 2 for a movie of **a**–**c** as increasing BE factor is applied.

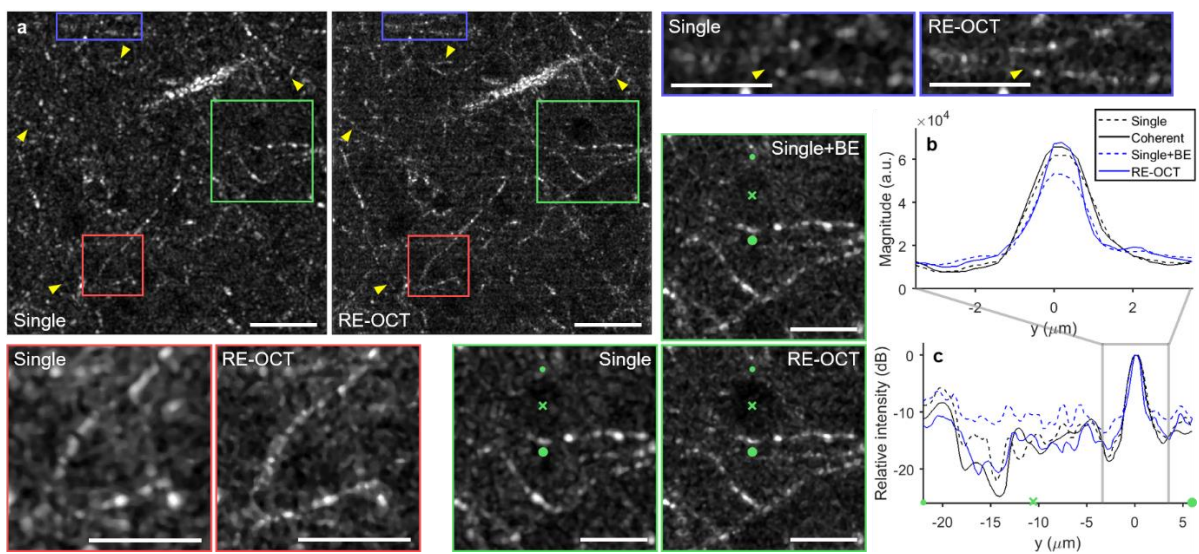

**Figure 5 RE-OCT with 100 acquisitions and 2.0× BE in the cortex of *ex vivo* mouse brain. a**, Single-shot and RE-OCT *en face* OCT images with zoomed insets regions indicated by boxes. Images were taken in the first cortical layer at approximately 100 µm below surface. Yellow arrows indicate myelinated axonal processed that can be more clearly visualized with RE-OCT. × markers in the green zoomed insets indicate one of the neurons, which appear as darker circles due to weak OCT scattering. Green inset shows that neuron in BE single-shot image was barely discernible due to the SNR penalty without coherent-average noise suppression. Scale bars, 40 µm (full) and 20 µm (zoomed). **b** and **c**, Cross-sectional profiles of a line connecting from small to large green dots in the green zoomed insets in a on linear and peak-normalized log scales. The green × marker indicates its corresponding position on the image. See Supplementary Movie 3 for a movie of **a**–**c** as increasing BE factor is applied.

In *ex vivo* fresh mouse brain, RE-OCT enhanced the visualization of myelinated axonal processes, especially the low-contrast features that were less apparent in the traditional single-shot image (Fig. 5a, yellow arrows). Although the narrowing of fibre width could be observed in the BE single-shot image, the peak signal magnitude was degraded without coherent-average noise suppression (Fig. 5b), similar to the effects in collagen gel. Due to the typical fibre thickness of 1-3 µm (which is comparable to the native OCT transverse resolution of 2.1 µm) of myelinated axons[52], some of the fibre narrowing observed here is not as prominent as in the collagen gel. The SNR penalty of the computational BE procedure is most apparent in the neuron (Fig. 5a, green inset), which produces lower OCT intensity than the surrounding brain tissue. Although the neuron remained visible in the RE-OCT image, the noise level in the BE single-shot image was brought up to that of the backscattered signal from the surrounding brain tissue, causing the neuron to 'disappear' into the background (Fig. 5a,c). This emphasizes the importance of coherent averaging in RE-OCT, particularly when weak-scattering structures need to be clearly visualized. Computational BE with BE factors larger than 2.0 resulted in degraded SBR and lower contrast between the neurons and surrounding brain tissue, without further narrowing of the fibre width or improvement to the peak signal magnitude (Supplementary Movie 3).

**Factors that limit achievable resolution enhancement**
We revisit the image signal power and phase in the spatial-frequency domain (Fig. 2c–e) to further understand the role of system noise, background, and aberrations on RE-OCT resolution. System noise limits not only the available DR of the spatial-frequency-domain image, but also the spatial-frequency bandwidth over which signal phase (associated with any given point scatterer in space) remains correlated (Fig. 2d,e and Supplementary Fig. 7). Phase correlation in the spatial-frequency domain has a direct implication on the spatial resolution of a coherent image—in order to achieve the best localization of signal energy in space, signal at different spatial frequencies must be able to constructively interfere (i.e., be in-phase with each other). Thus, the phase-correlation limit (which is limited by SNR in our experiment) determines how much of the expanded spatial-frequency bandwidth (determined by the BE factor) can support constructive interference and contribute to enhancing the resolution in RE-OCT. Computational BE far beyond the phase-correlation limit only serves to amplify the contribution of phase-decorrelated higher spatial frequencies, which degrades the SBR and the quality of the PSF without further improving the resolution (Fig. 3d–f and Supplementary Movie 1).

In contrast to system noise, background is composed of backscattered (single- (SS) and multiple-scattering (MS)) signal from the sample medium (silicone in this case). The spatial-frequency spectrum of the SS background is bandlimited and obeys the imaging bandwidth support of the system (determined by the illumination beam width in our system). Meanwhile, evidence has shown that frequency content of MS background may extend beyond the imaging bandwidth of the system[53]. In either case, background may exhibit uncorrelated phase as opposed to a flat profile of an ideal PSF (Supplementary Fig. 6b) and contribute to the disruption of phase correlation within (SS case) as well as outside (MS case) of the imaging bandwidth. As a result, resolution may be degraded by the presence of background compared to if the medium were completely transparent. In this respect, the role of background on OCT resolution is similar to that of optical aberrations—while aberrations contribute slowly varying phase inside the pupil, background contributes uncorrelated phase that results in the OCT speckle. Importantly, both effects imply that the sample itself may limit the achievable resolution; there can be contribution from sample-induced aberrations in addition to system aberrations, and the degradation of resolution by uncorrelated background phase becomes more severe when the structure of interest has lower SBR (e.g., due to weak scattering from the structure or strong scattering from the medium, or both). Furthermore, both background and aberrations are factors that cannot be mitigated by coherent-average noise suppression.

**Expanded framework of information capacity and resolution in coherent imaging**
A fundamental limit to resolution enhancement by RE-OCT is governed by the disruption of phase correlation in the spatial-frequency domain—due to system noise, background, aberrations, and other factors (e.g., sample instability, mechanical vibration, etc.). Among other factors, system noise played the most significant role in our experiments by determining the available DR of the image and the phase-correlation limit in the spatial-frequency domain. Notably, system noise is also the only factor that can be suppressed via coherent averaging in our experiments. Thus, the basis of RE-OCT lies in navigating the trade-off between resolution (in the space domain) and DR (in the spatial-frequency domain) of the image via coherent-average noise suppression and computational BE (Fig. 1a), where DR represents the impact on SNR that manifests in the spatial-frequency domain (Fig. 2d). In Fig. 3a–c, we prioritized resolution enhancement and applied a BE factor of 2.4, which sacrificed more DR than the 20 dB earned with coherent averaging (Supplementary Fig.8b). Alternatively, we could apply a BE factor of only 1.4 and simultaneously improve both resolution and SBR by a smaller margin (Fig. 3d,e), where the SNR penalty was offset by the increased peak PSF intensity as a by-product of improved localization of the PSF in space (note the maxima in Fig. 3e).

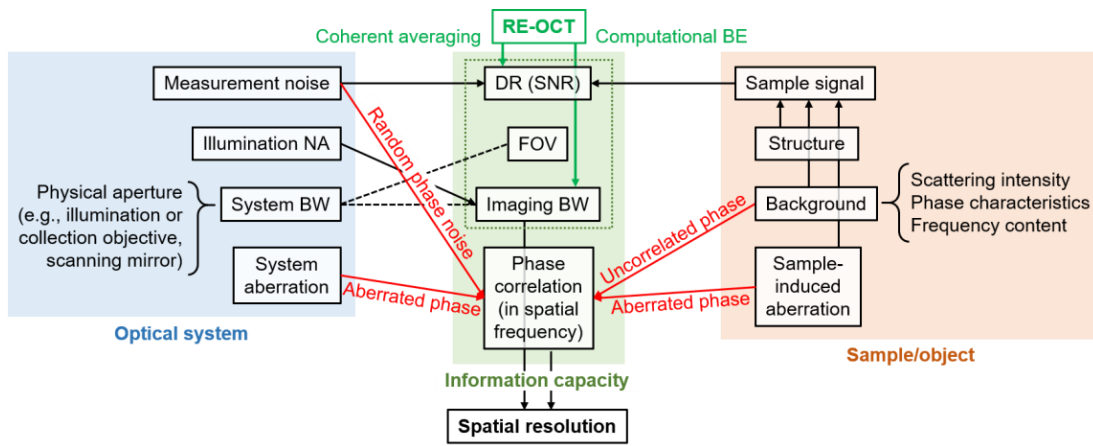

**Figure 6 Expanded framework of information capacity and resolution in coherent imaging.** Factors that influence elements of the expanded information capacity and ultimately affect resolution in coherent imaging. Phase correlation in the spatial-frequency domain is an important addition to the existing framework of information capacity (green dotted box)[4]. Red arrows denote factors that can disrupt phase correlation in the spatial-frequency domain. Green arrows indicate the role of RE-OCT: enhancing DR in the spatial frequency domain via coherent-average noise suppression, then, expanding imaging bandwidth via computation BE. For the sake of simplicity, the depicted framework omits the temporal components and only considers information in the spatial dimensions. Supplementary Section VII discusses the relationship between system and imaging bandwidths, illumination NA, and FOV. BW, bandwidth.

In order to reconcile the predictions of information capacity and our experimental RE-OCT results, we propose an expanded framework of information capacity and resolution in coherent imaging (Fig. 6). The expanded framework emphasizes phase correlation in the spatial-frequency domain (in addition to SNR, FOV and spatial-frequency bandwidth in Cox and Sheppard's framework[4]) as an important facet of the information capacity of a coherent imaging system. In theory, resolution is governed by the imaging bandwidth of the optical system. In practice, however, phase-correlation limit in the spatial-frequency domain (Fig. 2e and Supplementary Fig. 7) must also be considered when determining the best achievable resolution. Supplementary Section VIII computes the resolution enhancement that is theoretically supported by Cox and Sheppard's information capacity framework[4], and exemplifies the additional practical limit imposed by the SNR-limited phase-correlation limit (Supplementary Fig. 8a). Importantly, any factors—whether associated with the optical system or the sample itself—that can disrupt phase correlation in the spatial-frequency domain may prevent the optimal resolution (determined by the bandwidth support of the optical system) from being experimentally realized. By extension, in a time-dependent system whose information capacity includes the time-bandwidth product[4], the temporal resolution of such a system would also be subjected to the correlation of phase in the temporal-frequency domain.

**Discussion**

Beam-scanning is by far the most common mode of acquisition in OCT. Most beam-scanned OCT systems utilize a Gaussian illumination beam that under-fills the physical aperture limit (e.g., objective lens) in the optical system. This under-filling is required for the widely used telecentric scanning scheme, which is implemented to minimize distortions due to coherence gate curvature. (Although ophthalmic OCT systems do not implement telecentric scanning, the typical beam diameter of 1.2 mm in standard clinical systems[54] still under-fills the physical aperture of typical pupil diameters of adult eyes[55]). RE-OCT exploits this ubiquitous design feature in OCT systems to enable resolution enhancement beyond the aberration-free limit achieved by existing aberration correction approaches. In beam-scanned OCT, the aberration-free resolution limit is determined by a combination of the illumination beam width and the objective NA (Supplementary Section II); higher resolution is traditionally achieved by increasing the beam width or simply switching to a higher-NA objective (a standard practice in optical microscopy). RE-OCT, on the other hand, offers the flexibility to enhance transverse resolution beyond the aberration-free limit, using a simple computational procedure, without requiring any modification to the optical system or specialized acquisition schemes in existing aperture synthesis techniques. This gives RE-OCT a remarkable potential to have a widespread impact as it can be readily implemented on most OCT systems in the world and immediately unlock information that is beyond the current imaging capability of the system.

Phase stability is vital for achieving optimal performance in RE-OCT, as is the case in other phase-sensitive techniques in coherent imaging such as CAO and OCT angiography, which poses potential challenges for biological (e.g., live-cell imaging) and clinical (e.g., *in vivo* imaging) applications. In the case of RE-OCT, efficient earning of SNR via coherent-average noise suppression is contingent upon the backscattered signal from multiple acquisitions being phase-registered to each other. Our experiments in *ex vivo* fresh mouse brain (Fig. 5) required an additional image registration procedure (Supplementary Section V) to achieve phase registration before computing the coherent average. Others in OCT have also developed image registration and phase correction methods that successfully enabled phase-sensitive processing in biological samples, including for *in vivo* settings[22,56-58]. Furthermore, advances in high-speed imaging have enabled OCT imaging at MHz-rate[59], combatting motion artefacts and supporting phase-sensitive imaging *in vivo*[24-26,46,60-62]. RE-OCT may be particularly attractive for high-speed systems, some of which already incorporate coherent or incoherent averaging[46,60,61], where the RE-OCT procedure can be easily integrated into the existing imaging workflow with minimal additional effort and imaging time. Thus, RE-OCT can draw from existing and emerging solutions in the field to address the challenges associated with achieving the required phase stability for biological and clinical applications, including *in vivo* imaging.

Beyond the potential applicability of RE-OCT for OCT systems worldwide, the concept that noise suppression (improved SNR) can be harnessed for resolution enhancement—where efficient noise suppression is key to 'purchasing' greater resolution enhancement—has broad implications for not only OCT but also optical microscopy and imaging science. Traditionally, averaging and noise suppression have been associated with the improvement of image contrast or SBR in optical microscopy. However, the *theorem of invariance of information capacity* suggests that there are opportunities to exploit the information gained via noise suppression in other facets of optical imaging. RE-OCT applies this concept to improve resolution through an exchange of information between spatial-frequency bandwidth and SNR.

Furthermore, the expanded framework of information capacity and resolution in coherent imaging presented here is broadly relevant and contributes to the theory governing coherent image formation. This includes coherent imaging with an aperture-filled system (e.g., full-field OCT) even though RE-OCT is only applicable when the physical aperture is under-filled. For instance, an image of a sample that generates particularly weak signal may have a very limited DR in the spatial-frequency domain, such that the SNR-limited phase-correlation limit is *smaller* than the imaging bandwidth of the optical system. Resolution would be limited by the available DR as opposed to the illumination NA (in an under-filled system) or the objective NA (in an aperture-filled system) in such a low-SNR scenario. Alternatively, the detected backscattered signal may be well above the noise floor, but both the object and the surrounding medium contribute comparable signal strength such that the DR spanned by the SS signal level in the sample is low. The disruption of phase correlation by the SS and MS background could limit the ability to resolve the object in such a low-DR scenario. By extension, physically increasing the image bandwidth of an optical system (e.g., by using a higher-NA objective or increasing the illumination beam width) would yield the optimal improvement in resolution *only if* the acquired image had sufficient DR in the spatial-frequency domain to support phase correlation over the increased bandwidth. Thus, our expanded framework highlights important practical considerations (associated with both the optical system and the sample/object) for resolution in all forms of coherent imaging.

Future development may combine RE-OCT with aberration-diverse OCT[49] in order to suppress both the system noise and the MS background. Additionally, RE-OCT may be advantageous for imaging transversely isotropic structures (i.e., spatially *invariant* along a given spatial dimension) such as aligned muscle fibres or organized collagen fibrils in tendon and cartilage. This could allow for bandwidth along both the temporal *and* the invariant spatial dimension to be sacrificed to further enhance the resolution along the orthogonal spatial dimension[3].

**Conclusion**
RE-OCT is an approach that offers the flexibility to enhance resolution in beam-scanned OCT beyond the aberration-free resolution limit of the optical system. RE-OCT can be readily implemented on most OCT systems in the world without requiring any modification to the optical system. Based on the *theorem of invariance of information capacity*, RE-OCT navigates the information exchange between resolution and SNR by 'earning' SNR via coherent-average noise suppression, in order to 'purchase' superior resolution via computational BE. Coherent averaging was shown to increase DR in the transverse spatial-frequency domain, and for the first time, has been harnessed for resolution enhancement in OCT. In silicone phantom, RE-OCT achieved a resolution improvement of 1.5× (NA of 0.2 to 0.3), while maintaining comparable SBR to the traditional single-shot image. In collagen gel and *ex vivo* mouse brain, RE-OCT significantly enhanced the visualization of fine microstructural features, including low-contrast features that were otherwise obscured in the traditional OCT image. We found that the phase-correlation limit represents an additional practical limit to the effective spatial-frequency bandwidth support of a coherent imaging system that can be more restrictive that the theoretical limit imposed by the existing theory of information capacity[4]. Based upon these insights, we presented an expanded framework of information capacity and resolution in coherent imaging to incorporate these factors. This framework emphasizes the fundamental role of phase correlation, which contributes important implications to the theory of coherent imaging.

## Methods
### Optical system
The optical system was a standard telecentric beam-scanned spectral-domain (SD)-OCT system (Supplementary Fig. 2a). The SD-OCT system was sourced by a broadband superluminescent diode with a central wavelength of 850 nm and a bandwidth of 120 nm (Superlum, M-T-850-HP-I). Spectral data was detected by a spectrometer with a bandwidth of 180 nm (Wasatch Photonics, Cobra 800) and a 2048-pixel line-scan camera (e2v, Octopus). The sample arm utilized a double-pass illumination/collection configuration with an inverted 20× microscope objective with an NA of 0.45 (Olympus, LCPLN20XIR). Telecentric beam-scanning was accomplished with a 2-axis galvanometer and a zero-magnification telescope, which imaged the galvanometer to the back focal plane of the objective. The illumination beam diameter was ~2-3 mm, which under-filled the objective back aperture diameter of 8.1 mm. The native transverse resolution was 2.1 µm at the focal plane and the axial resolution was 1.9 µm in air. The system sensitivity was ~90 dB at the implemented acquisition rate (see RE-OCT image acquisition procedure) with a fall-off of –5 dB/mm. The system was controlled by a custom-built LabVIEW acquisition software.

### Sample preparation
All samples were prepared in glass coverslip-bottomed petri dishes, where the OCT beam interrogated the sample from the bottom through the coverslip (Supplementary Fig. 2c). The "noise image" for measuring the system noise in Fig. 2b,d was acquired by imaging the empty sample dish (Supplementary Fig. 2c).

Silicone phantom (Figs. 1 and 2) was prepared with a mixture of polydimethylsiloxane (PDMS) fluid (Clearco Product, PSF-50cst) and 2-part RTV silicone (Momentive Performance Materials, RTV-615 CLEAR 1#) at a weight ratio of 100:10:1 PDMS to RTV A to RTV B. Titanium dioxide particles with diameter of 0.5 µm were dispersed as scattering particles. Silicone mixture was baked at 70 °C for at least 8 hours to complete the polymerization process. The sample was stored room temperature, where the temperature was allowed to stabilize, prior to imaging.

Collagen gel (Fig. 4) was prepared with type I collagen (Corning, Collagen I, rat tail) at a final collagen concentration of 2.0 mg/mL. Collagen was polymerized at 4 °C for 15 min., 20 °C for 15 min., and finally 37 °C for 15 min. to promote formation of heterogeneous fibre architecture with thick collagen fibres[63]. The sample was removed from incubation 1-2 hours before imaging to allow the temperature to stabilize at room temperature.

*Ex vivo* mouse brain (Fig. 5) was harvested from a C57BL/6 mouse following euthanasia and perfusion with phosphate-buffered saline (PBS) at 4 °C. The harvested brain was stored in PBS at 4 °C before embedded in 1% agarose (Sigma-Aldrich, Agarose, low gelling temperature) in the sample dish without fixation. The sample was kept at room temperature for 1-2 hours to allow the temperature to stabilize at room temperature before imaging.

### RE-OCT image acquisition procedure
Images were acquired in CM mode, where 3D OCT volumes were acquired successively to allow sufficient decorrelation of noise (see Supplementary Section III). Each volume was acquired with a line scan rate of 70 kHz, an exposure time of 10 µs, and a transverse spatial sampling of 0.4 µm/pixel. In order to maximize the dynamic range spanned by the signal from the sample, image was acquired in the *conjugated configuration* by adjusting the reference arm such that the coverslip-bottom of the sample dish was positioned at larger pixel depths near the bottom of the B-scan (see Supplementary Section II).

**RE-OCT image reconstruction procedure**

RE-OCT image reconstruction from the acquired CM-mode volumes followed the procedure described in Supplementary Section IV. Briefly, space-domain OCT volumes were obtained from the raw tomograms via standard OCT image reconstruction, then, corrected for defocus via computational image formation procedures based on previously described methods[64]. For *ex vivo* mouse brain, an additional image registration procedure was required to correct bulk sample shift and phase drift in order to ensure that backscattered signal was spatially- and phase-registered across CM-mode volumes, as described in Supplementary Section V. Then, coherent average across processed OCT volumes was computed and its magnitude spectrum was obtained from the 2D transverse Fourier transform. A BE mask was computed from the magnitude spectrum at a given BE factor and applied to the coherent-average OCT volume in the transverse spatial-frequency domain. Finally, the BE spectrum was zero-padded to upsample (in space) before returning to the space domain. The spatial upsampling was implemented to facilitate resolution measurement via curve-fitting to the PSF. All RE-OCT image reconstruction and subsequent image processing was performed in MATLAB R2017a. All OCT images shown has undergone defocus correction and represent the traditional aberration-free imaging capability before computational BE.

**Calculations of OCT intensity**

The OCT intensity of the scattering particles, silicone background, and noise in Fig. 2b were computed as follows. Scattering particle intensity was obtained from the 99$^{th}$ percentile of the OCT scattering intensity (i.e., square of OCT magnitude) of the space-domain image at the focal plane. Silicone background intensity was obtained from the median of the OCT scattering intensity of the particle-removed space-domain image at the focal plane. The scattering particles were removed from the *en face* image via magnitude thresholding followed by a dilatation of the binary mask. Noise intensity was obtained from the standard deviation of the OCT scattering intensity of the "noise image" at the same pixel depth as the focal plane of the silicone phantom image. The "noise image" was obtained by imaging an empty blank sample dish (Supplementary Fig. 2c), placing the coverslip at the same pixel depth as in the silicone phantom.

**Calculation of dynamic range and phase-correlation limit**

The dynamic range in Fig. 2d and Supplementary Fig. 7 was computed as follows. First, the noise power spectrum was obtained from the square of the magnitude of the 2D transverse Fourier transform of the "noise image" at the same pixel depth as the focal plane of the silicone phantom image. Then, the relative noise power was computed w.r.t. the signal power at DC (i.e., $k_r = 0$ rad/μm) of the silicone phantom image at the focal plane. DR value in decibels was obtained from the mean of the relative noise power spectrum (uniformly distributed) across the entire transverse spatial-frequency domain.

The phase-correlation limit in Fig. 2e and Supplementary Fig. 7 was computed as follows. First, a window of size 243 × 243 pixels centred on a single scattering particle was cropped from the silicone phantom image at the focal plane. Then, the phase spectrum of the PSF was obtained from the angle of the 2D transverse Fourier transform of the window-out region. Next, local standard deviation of the phase spectrum was computed over a sliding kernel of size 3 × 3 pixels to obtain the "phase-decorrelation spectrum". The "phase-decorrelation spectrum" was divided into spatial-frequency bins, ranging from $k_r = 0$ rad/μm to $k_r = 7.85$ rad/μm (the Nyquist limit) at bin width of 0.1 rad/μm. Phase-correlation limit was obtained from the $k_r$ value at the centre of the bin at which the mean of the "phase-decorrelation spectrum" exceeded 0.2 rad.

**Measurements of resolution and SBR**

Resolution and SBR values in Fig. 3a,d–f were obtained from the Gaussian curve-fit to the PSFs (i.e., scattering particles) located at the focal plane. First, maximum intensity projection across 3 pixel-depths about the focal plane was computed from the OCT magnitude image. Scattering particles at the focal plane with peak magnitude > $5 \times 10^4$ were manually identified and the PSF images (a window of size 73 × 73 pixels centred on each particle) were cropped out. Then, a total of 32 radial cross-sectional profiles of the PSF images (i.e., 1D PSF profiles at 32 different angular cross sections) were extracted for linear least-square curve fitting to a 1D Gaussian function. The fit parameters from the 32 cross-sectional profiles were averaged to obtain the peak magnitude and full width at half-maximum (FWHM) of each particle. At this stage, particles with FWHM > 2.4 μm measured from the coherent-average OCT volume (i.e., noise-suppressed but not bandwidth-expanded) were excluded for being either air bubbles or aggregates of multiple particles. A total of 11 particles remained after the exclusion.

Resolution was obtained directly from the mean FWHM of the 11 remaining particles. SBR was obtained from the mean "peak SBR" of the 11 particles. The "peak SBR" in decibels of each particle was computed from the ratio of the peak PSF intensity (square of peak magnitude from the Gaussian fits described above) to the silicone background intensity (computed as described in Calculations of OCT intensity).

**Data availability**
Data underlying the results presented in this paper are not publicly available at this time but may be obtained from the authors upon reasonable request.

## Acknowledgements

The authors would like to thank Dr Justin C. Luo and Dr David M. Small for preparing the collagen gel and harvesting the mouse brain specimen, respectively. This work is funded in part by National Institutes of Health (NIBIB-R01GM132823, Adie) and National Science Foundation (CAREER: CBET-1752405, Adie).

## Author Contributions

N.L. prepared samples, performed all imaging experiments, developed RE-OCT reconstruction procedure, performed all data processing and analysis, and wrote the manuscript. S.G.A. conceived the principle of RE-OCT and provided guidance throughout this study. All authors reviewed and edited the manuscript.

## Additional Information

See Supplementary Information for supporting content.

## Competing Interests

The authors declare no competing interests.

# Resolution-enhanced OCT and expanded framework of information capacity and resolution in coherent imaging: supplementary information

Nichaluk Leartprapun[1] and Steven G. Adie[1,*]

[1]Nancy E. and Peter C. Meinig School of Biomedical Engineering, Cornell University, Ithaca, New York 14853, USA
*Correspondence should be addressed to sga42@cornell.edu

## Contents

# Supplementary Section I: Information capacity and resolution enhancement via exchange of information

Cox and Sheppard derived an expression for the information capacity (in bits) of an optical system[1,2] (Eq. 1), which we restate here:

$$C = (2L_x B_x + 1)(2L_y B_y + 1)(2L_z B_z + 1)(2TB_T + 1)\log_2(1 + s/n), \tag{S1}$$

where $L$, $T$, and $B$ denote the spatial field-of-view (FOV), temporal duration, and bandwidth in the associated dimension, respectively. The first three terms represent the space-bandwidth product (SBP) along the three spatial dimensions. The fourth term represents the time-bandwidth product (TBP). The last term represents the signal-to-noise ratio (SNR) in the logarithmic scale, where $s$ and $n$ denote the average signal power and additive noise power, respectively. For an imaging configuration where SBP, TBP and SNR $\gg$ 1, the information capacity of a single acquired volume simplifies to:

$$C_{\text{single}} = (2L_x B_x)(2L_y B_y)(2L_z B_z)(2TB_T)\log_2(s/n_{\text{single}}). \tag{S2}$$

If the object is known *a priori* to be invariant in time, the object can be successively imaged $N$ times (increasing $T$ by a factor of $N$) and the OCT datasets coherently averaged (reducing $B_T$ by factor of $N$), which leaves the time-bandwidth product unchanged, but suppresses the system noise by a factor of $1/N$—essentially collapsing the temporal dimension to enhance SNR. This process increases the information capacity of the coherent-average volume relative to the single-shot volume via the SNR term according to:

$$C_{\text{avg}} = (2L_x B_x)(2L_y B_y)(2L_z B_z)(2TB_T)\log_2(s/n_{\text{avg}}) \; ; \; n_{\text{avg}} = n_{\text{single}}/N \tag{S3}$$

$$\frac{C_{\text{avg}}}{C_{\text{single}}} = \frac{\log_2(s/n_{\text{avg}})}{\log_2(s/n_{\text{single}})} = 1 + \frac{\log_2(N)}{\log_2(s/n_{\text{single}})} \tag{S4}$$

Based on the *theorem of invariance of information capacity*, the extra information capacity earned via coherent-average noise suppression can be distributed to the SBP terms (e.g., by equally increasing $B_x$ and $B_y$) in order to enhance the resolution along those dimensions. According to the information capacity $C_{\text{avg}}$ given in Eq. S3, the allowable bandwidth expansion (BE) factor, $\varepsilon_{\text{BE}}$, that can be achieved without penalty to the SNR w.r.t. the original single-shot volume is given by:

$$\varepsilon_{\text{BE}}^2 = \frac{B_{x,\text{RE}} B_{y,\text{RE}}}{B_x B_y} = \frac{\log_2(s/n_{\text{avg}})}{\log_2(s/n_{\text{single}})}, \tag{S5}$$

where $B_{x,\text{RE}}$, and $B_{y,\text{RE}}$ are the expanded spatial-frequency bandwidths in the $x$ and $y$ dimensions, respectively. Note that it is the *square* of BE factor that scales with the factor of log-scale SNR gain (RHS of Eq. S5) because the extra SNR is equally distributed between the *two* spatial dimensions in this scenario. The resolution enhancement (RE) factor, $\varepsilon_{\text{RE}}$, supported by this information exchange process is then given by:

$$\varepsilon_{\text{RE}} = \frac{\text{Res}}{\text{Res}_{\text{RE}}} = \frac{B_{x,\text{RE}}}{B_x} = \frac{B_{y,\text{RE}}}{B_y} = \varepsilon_{\text{BE}}, \tag{S6}$$

where Res and $\text{Res}_{\text{RE}}$ denote the original and the enhanced transverse resolution, respectively. Note that since the $\varepsilon_{\text{RE}}$ is a function of the *ratio* of log-scale SNR, the base of the log does not influence the supported resolution enhancement (i.e., although log base 2 for SNR and information capacity in bits is used in Eq. S1–5, the same $\varepsilon_{\text{RE}}$ prediction could be obtained with log base 10 for SNR in dB scale). The factor of log-scale SNR gain and the corresponding supported RE factor are shown as a function of $N$ in Fig. S1. Supplementary Section VIII deals with more general cases where an arbitrary amount of SNR, not limited to the amount earned via coherent average, may be sacrificed to enhance resolution, as done in the silicone phantom experiment (Fig. 2).

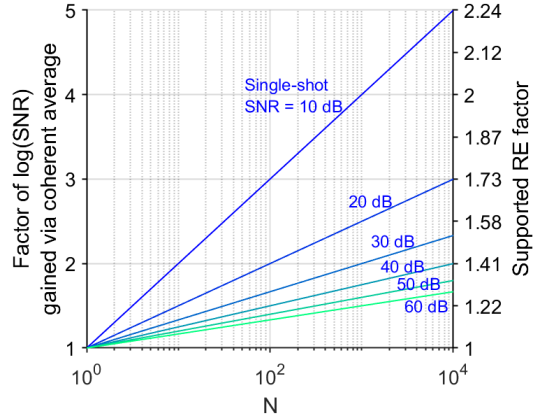

**Figure S1 Resolution enhancement supported by coherent-average noise suppression.** Gain in logarithmic-scale SNR and the corresponding RE factor supported by the invariance of information capacity as a function of $N$. Curves are shown for different original SNR of the single-shot volume. See Supplementary Section VIII for general cases where more SNR than earned may be sacrificed to enhance resolution.

## Supplementary Section II: System diagram, conjugated imaging configuration, and telecentric scanning

The spectral-domain (SD)-OCT system diagram is shown in Fig. S2a. Details of the optical components are provided in Methods. Imaging was performed in an inverted setup where the OCT beam interrogated the sample through the coverslip-bottom of the sample dish. The coverslip is essential for phase registration of individual OCT volumes (see Supplementary Section IV).

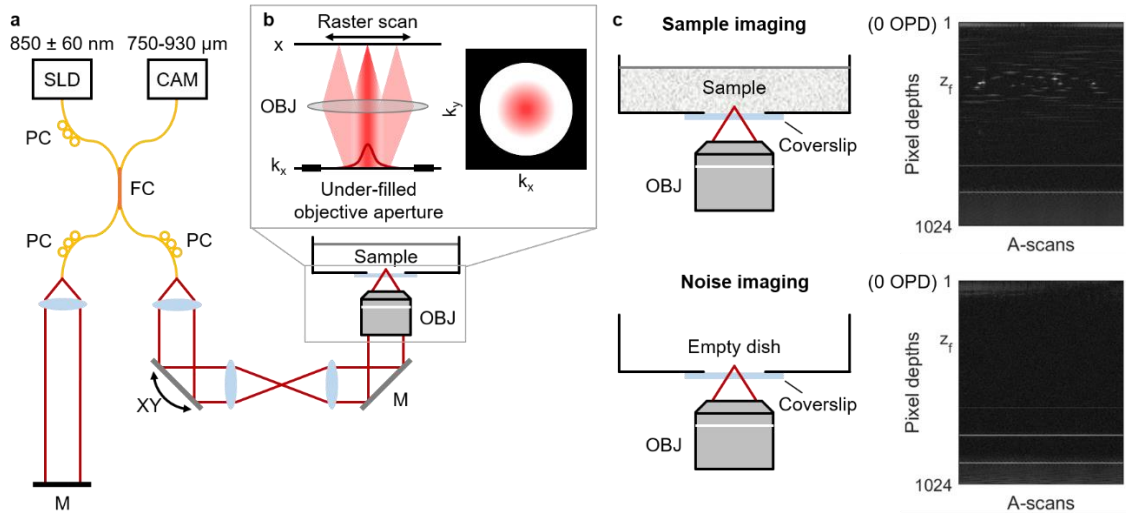

**Figure S2 Experimental setup. a**, SD-OCT system diagram. SLD, superluminescent diode. CAM, spectrometer and line-scan camera. PC, polarization controller. FC, 50/50 fiber coupler. M, mirror. XY, 2-axis galvanometer mirrors. OBJ, objective lens. **b**, Under-filling of objective aperture for telecentric scanning. **c**, Sample configuration and example cross-sectional OCT image demonstrating imaging in the conjugated configuration. $z_f$ indicates the focal plane of the sample image.

In this setup, the coverslip would appear closer to the 0 optical path difference (OPD) if it were imaged with the reference mirror positioned for imaging in the *traditional imaging configuration*, causing the signal at the camera to be dominated by the strong reflection at the glass-air and glass-sample interfaces of the coverslip. In other words, the actual signal of interest from the sample would only be able to occupy a small portion of the available dynamic range of the camera as the reference arm power was adjusted to avoid saturation at the coverslip interfaces. In order to maximize the dynamic range coverage of the signal from the sample, imaging was actually performed in the *conjugated configuration* where the reference mirror position was adjusted to wrap the coverslip toward higher pixel depths and placed the focal plane inside the sample closer to the 0 OPD (Fig. S2c). This configuration exploited the spectrometer roll-off to maximize signal from the focal plane while suppressing strong reflection from the coverslip.

The SD-OCT system in Fig. S2a utilized telecentric beam-scanning to acquire 3D tomograms. Telecentric scanning is the preferred beam-scanning scheme in OCT in order to minimize coherence gate curvature. In such a system, the objective aperture is underfilled by the illumination beam in order to allow for telecentric scanning (Fig. S2b). Thus, the numerical aperture (NA) of the system is determined by the width of the illumination beam (typically the $1/e^2$ width) rather than the objective NA, which sets a physical limit to the system bandwidth. The transverse resolution of the system can often be adjusted (to a certain extent) by changing the width of the illumination beam (by modifying the collimating or telescope optics) without switching the objective lens. However, there is a tradeoff between beam width and the FOV that can be supported without clipping the beam. Alternatively, an objective lens with higher NA can be used to achieve better transverse resolution. However, there is also a tradeoff between the system bandwidth (i.e., objective NA) and the supported FOV and working distance in a typical microscope.

## Supplementary Section III: Signal averaging in the space domain and caveats for achieving theoretical noise suppression performance

Figure 1b shows the signal intensity of the scattering particles, silicone background, and noise after incoherent and coherent average over N acquisitions. Signal intensity (i.e., OCT magnitude-square) of the scattering particles remained unchanged in both incoherent- and coherent-average images (Fig. 1b, triangle), indicating that the signal was dominated by phase-stable backscattering from the particles. In contrast, signal intensity in the silicone background was reduced by coherent average before stabilizing after $N > 10$ acquisitions (Fig. 1b, asterisk), suggesting that the silicone medium generated phase-stable in time (albeit low-magnitude) backscattering signal that was initially 'hidden' by noise and later revealed by coherent average. Faint scattering signal from the silicone background can be at the focal plane of the coherent-average, but not the traditional single-shot image (Fig. 1a, cross-sectional image). Meanwhile, the incoherent average failed to suppress the average background intensity (Fig. 1a,b). Noise images of an empty sample dish were acquired to quantify the system noise (Supplementary Fig. 2c), defined as the standard deviation of signal intensity across the transverse FOV. Both incoherent and coherent averages suppressed the noise, but the coherent average was more efficient (Fig. 1b, circle). Noise reduction followed the theoretical trends of $1/\sqrt{N}$ and $1/N$ for incoherent and coherent averages, respectively[3,4], demonstrating a factor of $\sqrt{N}$ superior efficiency in noise suppression by coherent over incoherent average.

The theoretical coherent-average noise suppression efficiency with a factor of $1/N$ under is based on the premise that different realizations of noise, a circularly symmetric complex random variable, are uncorrelated. For the system employed in this study, this condition was achieved when successive OCT volumes, acquired with a CM-mode acquisition scheme (see Methods), were coherently averaged (Fig. 1b and Fig. S3a, solid lines). However, noise suppression failed to reach the theoretical $1/N$ efficiency when successive B-scans, acquired with a BM-mode acquisition scheme (i.e., acquiring multiple B-scans at a slow-axis position, then, step to the next slow-axis position), were coherently averaged (Fig. S3a, dashed lines). These results suggest that acquisition of multiple B-scans was required to ensure decorrelation between different realizations of noise being averaged. Notably, when nonconsecutive B-scans from the same BM-mode datasets were coherently averaged (i.e., averaging every $\Delta T > 1$ B-scans), the noise suppression performance approached the theoretical efficiency with increasing $\Delta T$ (Fig. S3a, light dashed lines).

The distribution of noise power in the spatial-frequency domain provides additional insights into the results in Fig. S3a. A circular Gaussian noise is expected to be uniformly distributed across spatial frequencies. This was observed in the single-shot images acquired with both CM- and BM-mode acquisition schemes (Fig. S3b, left). The suppressed noise remained uniformly distributed after coherent average of successive OCT volumes (CM-mode), but not successive B-scans (BM-mode) (Fig. S3b, right), where noise was suppressed more efficiently at higher spatial frequencies. This result suggests that only the rapidly changing noise at higher spatial frequencies became decorrelated between successive B-scans.

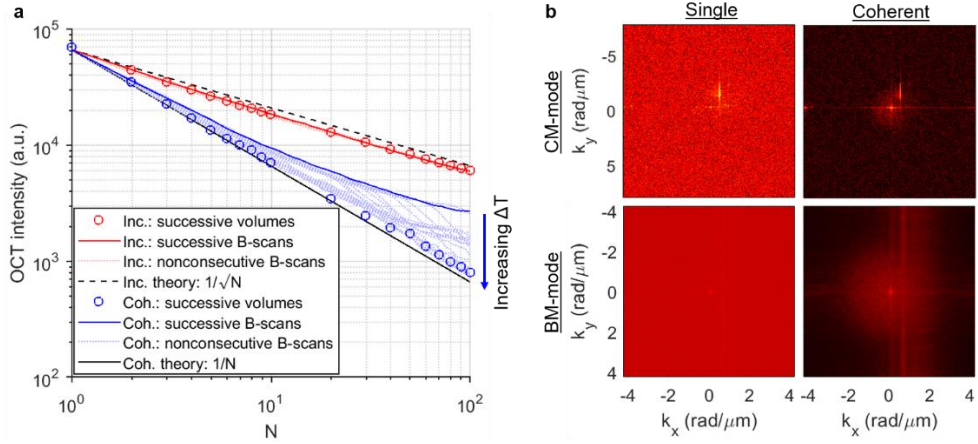

**Figure S3 Comparison of averaging schemes for noise suppression. a**, Noise reduction as a function of *N* for successive-volumes (circle), successive-B-scans (solid), and nonconsecutive-B-scans (light dotted) averaging schemes for incoherent (red) and coherent (blue) average. Coherent-average noise reduction with nonconsecutive B-scans approached the optimal efficiency of successive-volumes averaging scheme as $\Delta T$ increased. **b**, Single-shot and coherent-average noise power in transverse spatial-frequency domain for CM- and BM-mode acquisition schemes. For **a** and **b**, noise data was obtained from an *en face* plane located inside a glass coverslip.

## Supplementary Section IV: RE-OCT reconstruction procedure

The RE-OCT reconstruction procedure implemented in all experiments is illustrated in a flowchart in Fig. S4. First, the space-domain OCT volume from each acquisition was reconstructed by standard procedure (background subtraction, spectrum resampling, dispersion correction, and inverse Fourier transform). Defocus correction was performed on each reconstructed volume via computational image formation procedure (phase registration, bulk demodulation, and computational adaptive optics) as previously described[5]. As a result, each OCT volume has depth-invariant transverse spatial-frequency bandwidth and resolution. For *ex vivo* mouse brain, additional image registration procedure was necessary in order to ensure that individual volumes were phase registered to each other in the presence of sample instability during CM-mode acquisitions (Supplementary Section V).

The phase-registered OCT volumes were coherently averaged to obtain the *coherent-average OCT volume* via:

$$\tilde{S}_{\text{avg}}(x, y, z) = \frac{1}{N} \sum_{i=1}^{N} \tilde{S}_i(x, y, z). \tag{S7}$$

Then, the computational bandwidth expansion (BE) procedure began with computing of the *Fourier-domain OCT volume* via:

$$\tilde{S}_{\text{FD,avg}}(k_x, k_y, z) = \text{FT}_{x,y}\left[\tilde{S}_{\text{avg}}(x, y, z)\right], \tag{S8}$$

where $\text{FT}_{x,y}$ denotes 2D Fourier transform along the $x$ and $y$ dimensions, followed by a magnitude-average across depths about the focal plane $z_f$ to obtain the *magnitude spectrum*:

$$M(k_x, k_y) = \frac{1}{21} \sum_{i=-10}^{10} \left|\tilde{S}_{\text{FD,avg}}(k_x, k_y, z_{f+i})\right|. \tag{S9}$$

The depth averaging served to minimize the rapid noisy fluctuations in the magnitude spectrum that are present at a single plane.

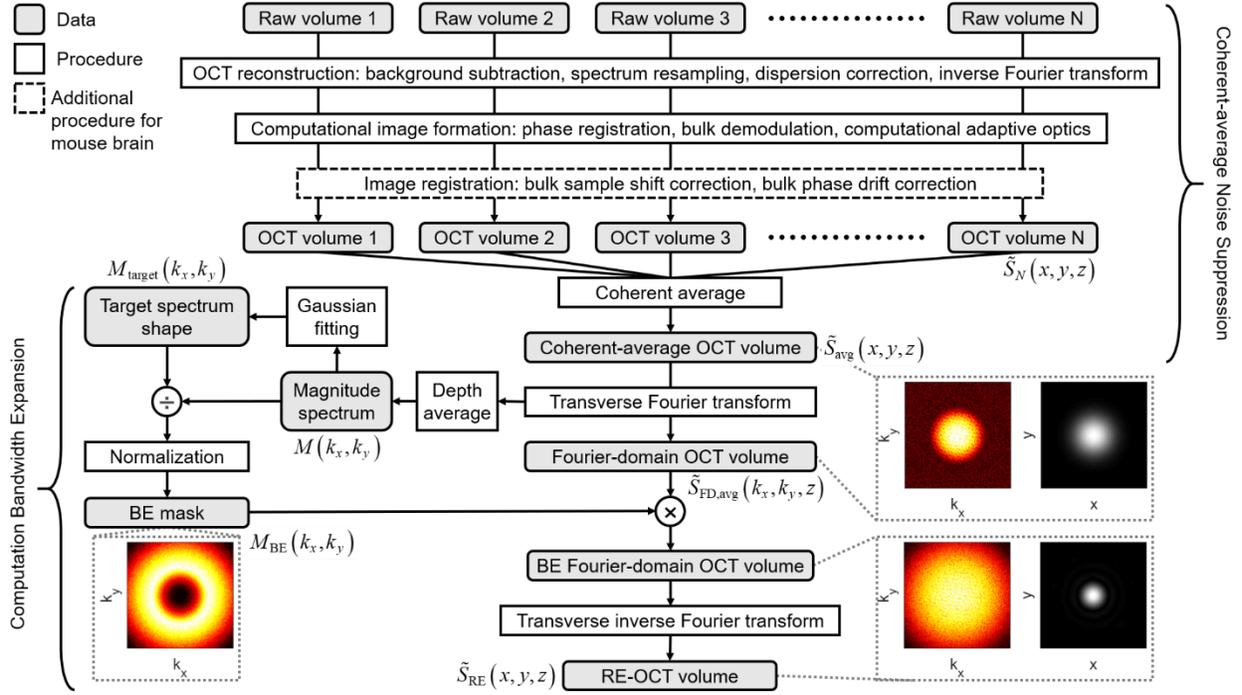

**Figure S4 RE-OCT reconstruction procedure.** Individual space-domain OCT volumes were reconstructed and processed for coherent-average noise suppression. Computational BE was performed on the coherent-average OCT volume via magnitude-based deconvolution to obtained the resolution-enhanced RE-OCT volume.

The depth-average *magnitude spectrum* was fit to a Gaussian curve as a function of the radial spatial frequency, $k_r = \sqrt{k_x^2 + k_y^2}$, then, the fit parameters $a$, $c$, and $d$ were used to compute the bandwidth-expanded *target spectrum shape* with a given BE factor, $\varepsilon_{BE}$, via:

$$M_{fit}(k_r) = a\exp(-k_r^2/c^2) + d, \tag{S10}$$

$$M_{target}(k_x, k_y) = a\exp\left(-\left(k_x^2 + k_y^2\right)/(\varepsilon_{BE}c)^2\right) + d. \tag{S11}$$

The *BE mask* was computed from the *target spectrum shape* and the *magnitude spectrum*, then, normalized to ensure that the total signal power would be conserved after computational BE, as follows:

$$M_{BE}(k_x, k_y) = \frac{1}{\varepsilon_{BE}^2} \frac{M_{target}(k_x, k_y)}{M(k_x, k_y)}. \tag{S12}$$

Finally, the resolution-enhanced RE-OCT volume was obtained via magnitude-based deconvolution in the spatial-frequency domain:

$$\tilde{S}_{RE}(x, y, z) = FT^{-1}_{x,y}\left[M_{BE}(k_x, k_y)\tilde{S}_{FD,avg}(x, y, z)\right], \tag{S13}$$

where $FT^{-1}_{x,y}$ denotes 2D inverse Fourier transform along the $x$ and $y$ dimensions.

### Supplementary Section V: Image registration procedure for *ex vivo* mouse brain

The *ex vivo* mouse brain in this study experienced both bulk sample shift and bulk phase drift, likely due to the temperature stabilization of the cold mouse brain and the warm mounting agarose (see Methods). In such case, image registration procedure was required to ensure that the scattering signal from different OCT volumes are phase-registered to each other prior to computing the coherent average. First, the bulk sample shift in 3D space was corrected via a Fourier transform-based image translation registration algorithm[6]. Each OCT volume was conjugated to the 1st volume in the spatial-frequency domain, before its bulk spatial shifts relative to the 1st volume along $x$, $y$, and $z$ dimensions were estimated from the peak position of space-domain impulse response via:

$$\tilde{S}_{\text{FD},i}\left(k_x, k_y, k_z\right) = \text{FT}_{x,y,z}\left[\tilde{S}_i\left(x, y, z\right)\right] ; i = 1, 2, \ldots, N , \tag{S14}$$

$$(\Delta x_i, \Delta y_i, \Delta z_i) = \arg\max_{(x,y,z)}\left(\text{FT}^{-1}_{x,y,z}\left[\tilde{S}_{\text{FD},i}^*\left(k_x, k_y, k_z\right)\tilde{S}_{\text{FD},1}\left(k_x, k_y, k_z\right)\right]\right), \tag{S15}$$

where $\text{FT}_{x,y,z}$ and $\text{FT}^{-1}_{x,y,z}$ denote the 3D forward and inverse Fourier transform, and $\tilde{S}_{\text{FD},i}^*$ denotes the complex conjugate of $\tilde{S}_{\text{FD},i}$, respectively. Then, the bulk spatial shifts $\Delta x_i$, $\Delta y_i$, and $\Delta z_i$ were applied back to the $i^\text{th}$ volume as phase ramps to spatially register it to the 1st volume via:

$$\tilde{S}_{i,\text{shift-corrected}}\left(x, y, z\right) = \text{FT}^{-1}_{x,y,z}\left[\tilde{S}_{\text{FD},i}\left(k_x, k_y, k_z\right)\exp\left(-j2\pi\left(k_x\Delta x_i + k_y\Delta y_i + k_z\Delta z_i\right)\right)\right], \tag{S16}$$

where $j = \sqrt{-1}$.

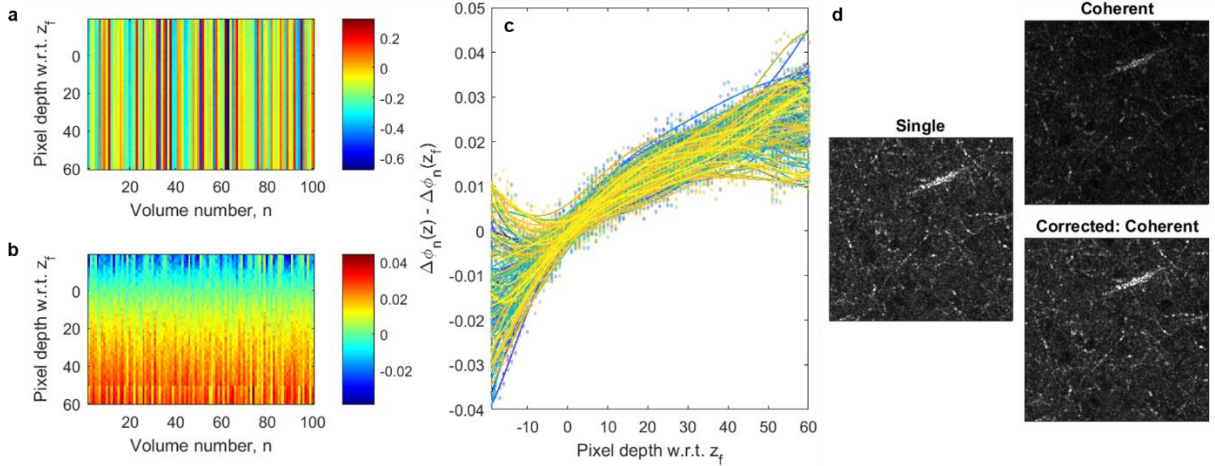

**Figure S5 Bulk phase drift correction in *ex vivo* mouse brain. a,** Bulk phase difference between adjacent volumes $\Delta\phi_i(z)$ at each pixel depth. **b,** Bulk phase difference in **a** w.r.t. the focal plane, $\Delta\phi_i(z) - \Delta\phi_i(z_f)$, to show the depth dependence. **c,** Depth-dependent polynomial curve fits of bulk phase difference for different volumes in **b**. **d,** Single-shot and coherent-average images with and without image registration (same colormap range).

After correcting for the bulk sample shift, a bulk phase difference between adjacent volumes was estimated at each pixel depth $k$ via:

$$\Delta\phi_i(z_k) = \angle\left(\sum_{x,y}\left[\tilde{S}_{i,\text{shift-corrected}}\left(x, y, z_k\right)\tilde{S}_{i-1,\text{shift-corrected}}^*\left(x, y, z_k\right)\right]\right) ; i = 2, 3, \ldots, N, , \tag{S17}$$

which represents the magnitude-weighted average phase difference across each *en face* plane (Fig. S5a). Each of the $i^\text{th}$ phase difference was fit to a 6$^\text{th}$-order polynomial function as a function of $z$ via linear least-square curve fitting to obtain a depth-dependent phase shift between adjacent volumes (Fig. S5c). The cumulative depth-dependent phase drift up to the $i^\text{th}$ volume was removed from each subsequent volume to phase-register it to the 1st volume via:

$$\tilde{S}_{i,\text{phase-registered}}(x, y, z) = \tilde{S}_{i,\text{shift-corrected}}(x, y, z) \exp\left(-j \sum_{n=1}^{i} \Delta\phi_n(z)\right); i = 2, 3, \ldots, N. \quad (S17)$$

Figure S5d shows the comparison between the coherent-average images of the *ex vivo* mouse brain with and without image registration.

## Supplementary Section VI: Simulation of RE-OCT in silicone phantom

In order to understand the factors that limit achievable resolution enhancement in RE-OCT, a set of simulated *en face* planes were generated with the same number of pixels (450×450 pixels), transverse FOV (180 µm × 180 µm) and spatial sampling (0.4 µm/pixel) as the silicone phantom datasets in Figs. 1 and 2. The simulated *en face* planes consisted of three components: scattering particles (Fig. S6a), silicone background (Fig. S6b), and system noise (Fig. S6c).

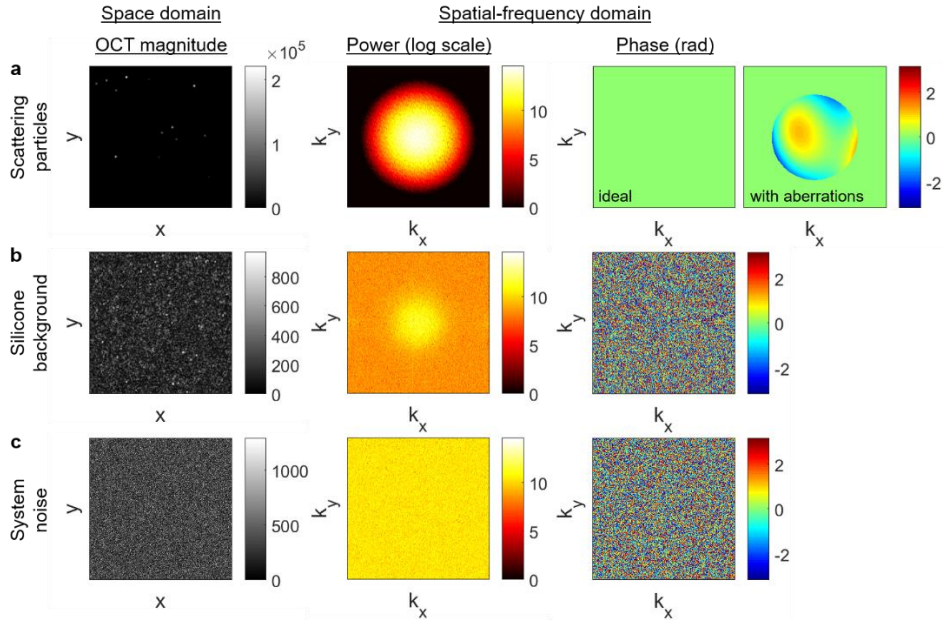

**Figure S6 Components of simulated *en face* planes. a–c**, Space-domain OCT magnitude image (left), and spatial-frequency-domain power (middle) and phase (right) spectra of simulated scattering particles, silicone background, and system noise, respectively. Phase spectra in **a** show the pupil phase applied to the simulated Gaussian PSFs. Note the power spectrum shape of the silicone background, in contrast to the uniformly distributed spectrum of the random system noise.

Each component was generated as follows:
- **Scattering particles.** 2D symmetric Gaussian point spread functions (PSF) with full width at half-maximum (FWHM) of 2.1 µm, matching the native transverse resolution of the system, and peak magnitudes matching the OCT signal magnitudes of the scattering particles from the focal plane of the silicone phantom datasets (Fig. S6a). For the simulated case with optical aberrations, an aberrated phase profile computed with Zernike polynomials were added to the Gaussian PSFs[7] (Fig. S6a, right).
- **Silicone background.** Complex OCT signal of the silicone medium from the focal plane of the silicone phantom datasets. First, the scattering particles were removed from the *en face* image via a magnitude threshold. Then, the gaps left behind at the particle locations were "filled in" by a patch of silicone image from another region (Fig. S6b).
- **System noise.** A 450×450 array of circularly symmetric complex random variable with a mean intensity (i.e., magnitude[2]) equivalent to that of the single-shot noise intensity from the focal plane of the silicone phantom datasets in Fig. 1b (Fig. S6c). For a coherent-average image across $N$ acquisitions, the array of complex random variable was repeatedly generated for $N$ iterations and coherently averaged.

A total of six simulated *en face* planes were generated: noise only (scattering particles + system noise), noise with background (scattering particles + silicone background + system noise), and noise-free limit (scattering particles only), each of the three cases with and without optical aberrations. The RE-OCT procedure was performed on each of the simulated *en face* planes as described in Supplementary Section IV. The resolution and SBR results were computed as described in Methods.

## Supplementary Section VII: Noise, dynamic range, and phase correlation in the spatial-frequency domain

System noise limits not only the available DR of the image, but also the spatial-frequency bandwidth over which signal phase remains correlated in spatial frequency. In order to investigate the results observed in Figs. 1c–e over a wider range of noise levels, simulated *en face* planes containing scattering particles and system noise (i.e., noise only case described in Supplementary Section VI) were generated for coherent average over a range of $N = 1$ (Fig. S7a) through $N = 10^5$ (Fig. S7b) acquisitions. DR and phase-correlation limit, $k_{\text{phase-corr}}$, in the spatial-frequency domain were computed for each simulated *en face* plane as described in Methods.

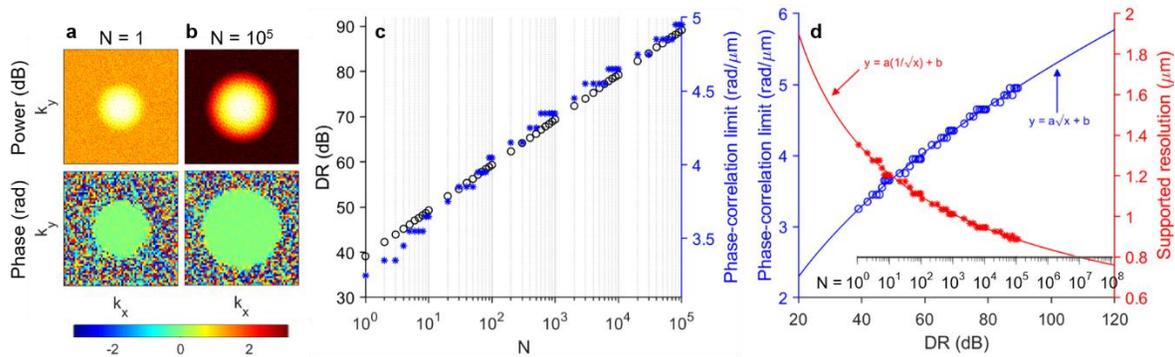

**Figure S7 Noise-limited dynamic range and phase-correlation limit. a** and **b**, Power and phase of simulated single-shot ($N = 1$) and coherent-average ($N = 10^5$) image in transverse spatial-frequency domain. **c**, DR (black) and phase-correlation limit (blue) as a function of $N$. **d**, Phase-correlation limit (blue) and the supported resolution (red) as a function of DR. Both simulation results (marker) and curve-fits (line) are shown. Inset shows corresponding $N$ required for the DR values.

Both DR and phase-correlation limit increases as the system noise is further suppressed via coherent average over larger $N$ (Fig. S7c). For the definition of the phase-correlation limit implemented here (see Methods), the phase-correlation limit corresponds to the spatial frequency at which the signal power is roughly 16 dB above the system noise floor. Given the quadratic drop of the Gaussian spectrum tail on the logarithmic scale, the increase in phase-correlation limit from coherent-average noise suppression becomes less efficient at increasing $N$, even though DR increases linearly with $N$ on the logarithmic scale (i.e., noise suppression factor of $1/N$). This diminishing efficiency can be seen on the plot of phase-correlation limit as a function of DR, where the phase-correlation limit scales with square root of DR (Fig. S7d, blue). The phase-correlation limit also provides an estimate of the best possible resolution that the system can support (Fig. S7d, red)—given by $\lambda/(2\text{NA}_{\text{max}})$ and $k\text{NA}_{\text{max}} = k_{\text{phase-corr}}$, where $k$ is the wave number—under the premise that signal from higher spatial frequencies beyond the phase-correlation limit cannot constructively interfere, thus, cannot contribute to the localization of PSF energy in the space domain.

## Supplementary Section VIII: Fundamental limits to resolution enhancement in RE-OCT

RE-OCT utilizes the framework for resolution enhancement based on gaining extra SNR via coherent-average noise suppression and the *theorem of invariance of information capacity* presented in Supplementary Section I. However, the amount of resolution enhancement in RE-OCT need not be limited to that of the SNR gain obtained through coherent averaging. In fact, the BE factor can be selected to achieve the most optimal combination of DR and resolution, depending on the SNR penalty that can be tolerated in an application. In a general case where any BE factor may be applied, the theoretically achieved RE factor and the accompanying SNR penalty based on Cox and Sheppard's information capacity framework[1] is:

$$\frac{\text{Res}_{\text{native}}}{\text{Res}_{\text{RE-OCT}}} = \frac{B_{\text{RE}}}{B_{\text{native}}} = \sqrt{\frac{\log_2(s/n_{\text{avg}})}{\log_2(s/n_{\text{RE}})}}, \tag{S19}$$

where $B_{\text{native}}$ and $B_{\text{RE}}$ denote the isotropic spatial-frequency bandwidth in both $x$ and $y$ dimensions before and after computational BE, respectively. In principle, the resolution is enhanced by as much as the bandwidth is expanded (i.e., RE factor = BE factor), and the bandwidth can be expanded by as much as one is willing to sacrifice the SNR (according to the *theorem of invariance of information capacity*). The SNR penalty (in dB) in exchange for a given BE factor is $10\log_{10}(n_{\text{avg}}/n_{\text{RE}})$, where $n_{\text{RE}} \geq n_{\text{avg}}$ for BE factor $\geq 1$. This is the minimum SNR that must be 'earned' via coherent average if one were to maintain the original single-shot SNR in the final RE-OCT image.

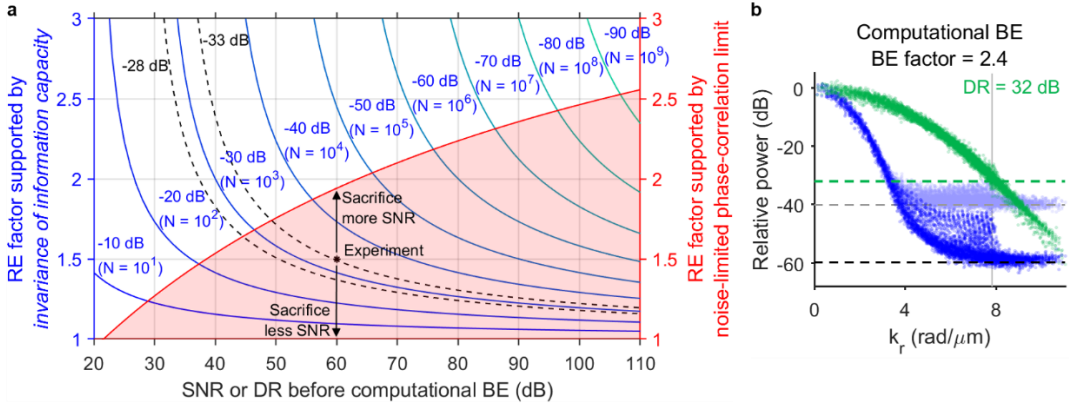

**Figure S8 Fundamental limits to resolution enhancement in RE-OCT. a**, RE factor as a function of initial SNR (or DR in the spatial-frequency domain) before computational BE supported by the *theorem of invariance of information capacity* at different amount of SNR penalty (blue) and the additional practical limit imposed by the noise-limited phase-correlation limit (red). Each blue curve is labeled with the SNR penalty and the number of acquisitions $N$ required in coherent-average noise suppression to fully compensate for the loss. Shaded region under the red curve represents the practical "RE-OCT operating range". Black marker indicates the experimental performance in silicone phantom in Figs. 1 and 2, where in principle, the performance may be adjusted up or down within the operating range to prioritize resolution or SNR, respectively. **b**, Relative power as a function of radial spatial frequency of the single-shot (light blue, DR = 40 dB), coherent-average (dark blue, DR = 60 dB), and RE-OCT (green, DR = 32 dB) images of the silicone phantom (Fig. 2a). An estimate of -28 dB in SNR has been sacrificed during the computational BE procedure to produce the RE-OCT image in Fig. 2.

The resolution enhancement supported by Cox and Sheppard's information capacity framework[1] with different amounts of SNR penalty is shown as a function of the initial SNR before computational BE (Fig. S8a, blue). Ironically, starting with a lower initial SNR supports larger RE factor for a given dB sacrificed (i.e., each individual curve has a decreasing trend). However, starting with a higher SNR means there is more SNR to sacrifice by computational BE before the final SNR of the bandwidth-expanded image drops below 0 dB. More importantly, this fundamental limit based on the *theorem of invariance of information capacity* does not account for the implication of noise on the phase correlation in the spatial-frequency domain—that is, the impact of phase decorrelation on disrupting the constructive interference that is required to produce the optimal resolution in space. The SNR-limited phase-correlation limit, $k_{\text{phase-corr}}$, in the spatial-frequency domain imposes another upper limit to the best possible resolution that the system can support, based on the spatial-frequency bandwidth over which phase remains correlated (Fig. S7d in Supplementary Section VII). Taking the simulation results in Fig. S7d and $\text{Res}_{\text{native}} = 2.1$ µm, the resolution enhancement supported by the noise-limited phase-correlation limit is shown (in red) on top of the theoretical information-capacity limit in Fig. S8a. Note that the simulation results in Fig. S7d only account for the effects of system noise; there are other factors that can disrupt phase correlation in the spatial-frequency domain, thus, further limiting the achievable resolution enhancement, in practice (as discussed in the main manuscript).

The red shaded region under the curve represents the practical "RE-OCT operating range". The experimental performance in silicone phantom (i.e., a coherent-average DR of 60 dB going into computational BE with a BE factor of 2.4, resulting in an RE factor of 1.5) is indicated by a black marker. In principle, this asterisk can be flexibly moved up (to achieve better resolution improvement while sacrificing more SNR) or down (to preserve more SNR while achieving less resolution improvement) within the shaded region, at a given initial coherent-average SNR going into computational BE (Fig. S8a, black arrows). The experimental performance suggests that roughly 33 dB in SNR was sacrificed during the computational BE procedure, where the -33 dB dashed curve intersects the black marker. This SNR penalty is slightly more than the -28 dB estimate based on the relative power spectrum of the bandwidth-expanded versus coherent-average image (Fig. S8b). These results suggest that there may be slight discrepancies between the true SNR in the space domain of the RE-OCT image (which we do not have an experimental measurement of) and the estimated DR in the spatial-frequency domain.

## Supplementary Movie 1 Caption
**Resolution-enhanced (RE)-OCT in silicone phantom with increasing bandwidth expansion (BE) factor. a**, Single-shot, coherent-average, BE single-shot, and RE-OCT power spectrums (log scale) and space-domain *en face* OCT image with zoomed PSF (linear scale). Resolution and SBR represent mean ± standard deviation of measurements from 11 particles. Scale bars, 40 µm (*en face* image) and 2 µm (zoomed PSF). **b** and **c**, Cross-sectional profiles of zoomed PSF in a on peak-normalized linear and log scales.

## Supplementary Movie 2 Caption
**Resolution-enhanced (RE)-OCT in fibrous collagen gel with increasing bandwidth expansion (BE) factor. a**, Single-shot and RE-OCT *en face* OCT images with zoomed insets regions indicated by boxes. Scale bars, 40 µm (full) and 20 µm (zoomed). **b** and **c**, Cross-sectional profiles of a line connecting from small to larger green dots in the green zoomed insets in a on linear and peak-normalized log scales.

## Supplementary Movie 3 Caption
**Resolution-enhanced (RE)-OCT in the cortex of *ex vivo* mouse brain with increasing bandwidth expansion (BE) factor. a**, Single-shot and RE-OCT *en face* OCT images with zoomed insets regions indicated by boxes. Images were taken in the first cortical layer at approximately 100 µm below surface. × markers in the green zoomed insets indicate one of the neurons, which appear as darker circles due to weak OCT scattering. Green inset shows that neuron in BE single-shot image was barely discernible due to the SNR penalty without coherent-average noise suppression. Scale bars, 40 µm (full) and 20 µm (zoomed). **b** and **c**, Cross-sectional profiles of a line connecting from small to large green dots in the green zoomed insets in a on linear and peak-normalized log scales. The green × marker indicates its corresponding position on the image.